\documentclass[fleqn,12pt,twoside]{article}
\usepackage{espcrc1}

\usepackage{graphicx}
\usepackage[figuresright]{rotating}

\newcommand{\AmS}{{\protect\the\textfont2
  A\kern-.1667em\lower.5ex\hbox{M}\kern-.125emS}}

\title{Baryon resonances from chiral coupled-channel dynamics}

\author{M.F.M.~Lutz\address[GSI]{Gesellschaft f\"ur Schwerionenforschung (GSI)\\
Planck Str. 1, 64291 Darmstadt, Germany },
E.E.~Kolomeitsev\address[UMN]{School of Physics and Astronomy, \\ University of Minnesota,
Minneapolis, MN 55455, USA}}

\begin{document}

\maketitle

\begin{abstract}
We discuss the formation of s- and d-wave baryon resonances as
predicted by the chiral SU(3) symmetry of QCD. Based on the
leading order term of the chiral Lagrangian a rich spectrum of
molecules is generated, owing to final-state interactions. The
spectrum of s- and d-wave baryon states with zero and non-zero
charm is remarkably consistent with the empirical pattern. In
particular, the recently announced $\Sigma_c(2800)$ of the BELLE
collaboration is reproduced with realistic mass and width
parameters. Similarly, the d-wave states $\Lambda_c(2625)$ and
$\Xi_c(2815)$ are explained naturally to be chiral excitations of
$J^P=\frac{3}{2}^+$ states. In the open-charm sector exotic
multiplet structures are predicted. These findings support a
radical conjecture: {\it meson and baryon resonances that do not
belong to the large-$N_c$ ground state of QCD should be viewed as
hadronic molecular states}.
\end{abstract}

\section{Introduction}

The task of constructing a systematic effective field theory for resonance physics of
QCD is one of the key challenges in hadronic physics. Considerable progress was achieved
in the last few years though there are certainly still some loose ends to be pondered over
at this stage \cite{Meissner:1999vr,LK01,LK02,LWF02,Ji01,grkl,Jido03,Granada,KL03}. Astounding
results have been worked out demonstrating that many meson and
baryon resonance states can be easily understood as being formed due to final
state interaction. The computations of the authors \cite{LK01,LK02,LWF02,Granada,KL03,LK04-axial,LK04-charm}
were driven by a hypothesis put forward
some years ago: meson and baryon resonances not belonging to the large-$N_c$ ground
states are generated dynamically by coupled-channel interactions \cite{LK01,LK02,LWF02}.
Upon selecting a few `fundamental' hadronic degrees of freedom the zoo of resonances is
conjectured to be a result of the interactions of the latter.
The identification of the proper set was guided by properties of QCD in the
large-$N_c$ limit. We do not support for instance the approach of Chew and
Low \cite{Chew-Low}, who viewed the isobar resonance, a member of the large-$N_c$ baryon
ground state of QCD, as being generated by pion-nucleon interactions. According
to \cite{LK01,LK02} the decuplet states should be considered on an equal footing with
the baryon octet states. Similarly, contrary to the work by Zachariasen and
Zemach \cite{Zachariasen-Zemach}, we consider the $\rho$-meson to be 'fundamental',
being a member of the large-$N_c$ meson ground states.

Most spectacular are results where the driving term of the final state
interaction is predicted unambiguously by the chiral symmetry of QCD. This is the
case for the s-wave interaction of the Goldstone bosons with any of the 'fundamental'
hadrons. Striking predictions for resonances with $J^P= 0^+,1^+,
\frac{1}{2}^-, \frac{3}{2}^-$ quantum numbers are the consequence. They are interpreted to be
chiral excitations of the large-$N_c$ ground states with $J^P= 0^-,1^-,
\frac{1}{2}^+, \frac{3}{2}^+$ quantum numbers. Though the
study of $0^+$ and $\frac{1}{2}^-$ molecules  \cite{Wyld,Dalitz,Torn,Rupp86}
has a long history, the claim that $1^+$ and $\frac{3}{2}^-$ resonances are molecules too
is novel and may be provocative to the community. In this talk we review the
formation of baryon resonance with charm quantum numbers zero and one.
New unpublished results on d-wave resonances with open charm  are included.
An interpretation of the newly discovered $\Sigma_c(2800)$
as a chiral molecule is given. Similarly, the previously established
d-wave states $\Lambda_c(2625)$ and $\Xi_c(2815)$ are reproduced naturally within
chiral coupled-channel dynamics.

\section{Chiral coupled-channel dynamics}

The basis of an effective field theory approach to the resonance spectrum of QCD is
the chiral Lagrangian \cite{book:Weinberg}. The latter manifests the chiral properties of
QCD in a systematic way, at the price of an infinite hierarchy of interaction terms. Many of
the vertices are predicted in terms of a chiral order parameter $f\simeq 90$ MeV, which parameterizes the
weak decay processes of the Goldstone bosons. Still, an infinite number of interaction terms is
unknown and has to be determined by either experiment or additional dynamical information from QCD,
like large-$N_c$ sum rules. Predictive power arises as a consequence of counting rules that
estimate the importance of a given contribution based on the naturalness assumption. The
important property is that at a given order of accuracy only a finite number of parameters enter
the calculation.

Most striking is the
leading-order prediction for the s-wave interaction of the Goldstone bosons, $\Phi_a(x)$,
with the baryon octet, $B_a(x)$, and decuplet fields, $B^\mu_{abc}(x)$.
Given the kinetic term of the baryon field the interaction follows by a chiral
rotation unambiguously in terms of the chiral order parameter:
\begin{eqnarray}
&&\mathcal{L}=
 \frac{i}{8\, f^2}\, {\rm tr}\, \Big((\bar B
\,\gamma^\mu\,\, B) \cdot
 [\Phi,(\partial_\mu \Phi)]_-  \Big)
+\frac{3\,i}{8\, f^2}\, {\rm tr}\, \Big((\bar B_\nu
\,\gamma^\mu\,\, B^\nu) \cdot
 [\Phi,(\partial_\mu \Phi)]_-  \Big)  \,.
 \label{WT-term}
\end{eqnarray}
The octet and decuplet fields, $\Phi, B$ and $B_\mu$, posses an appropriate matrix structure
according to their SU(3) tensor representation.

It is instructive to discuss first the SU(3) limit where the
scattering channels may be decomposed into invariants
\begin{eqnarray}
&& 8 \otimes 8 = 27 \oplus \overline{10} \oplus 10 \oplus 8 \oplus 8 \oplus 1 \,, \qquad \qquad
8 \otimes 10 = 35 \oplus 27 \oplus 10 \oplus 8\,.
\label{decomp}
\end{eqnarray}
According to (\ref{decomp}) the scattering of the Goldstone bosons of the baryon octet states
has 6 independent channels. The striking implication of QCD's chiral SU(3) symmetry is
that attraction is predicted in the two octet and singlet channels only, the 27plet channel is
repulsive. No interaction is provided in the decuplet channels. Amazingly, in
just the channels where attraction is foreseen a rich spectrum of baryon resonances is observed
in experiment. Those channels coincide with the prediction of the quark model which suggest that
baryon resonances are formed primarily in channels with quantum numbers which can be realized in
terms of three quarks only. A similar analysis reveals that the interaction of Goldstone bosons with
the decuplet states is strongly attractive in the octet and decuplet sectors, but repulsive in the
35plet. The 27plet sector shows weak attraction only. Relative to the octet channel its attraction
is reduced by a factor 6. The attraction in the decuplet channel relative to
the octet channel is reduced by the factor 2 only.

Interaction terms analogous to (\ref{WT-term}) can be written down describing the s-wave
interaction of Goldstone bosons with the open-charm $\frac{1}{2}^+$ and $\frac{3}{2}^+$ ground
states \cite{Wise92,YCCLLY92,BD92}. They form anti-triplet and sextet representations of the SU(3)
group. The Clebsch-Gordon coefficients that describe the relative interaction strength of the various channels reflect
the SU(3) multiplet structure of the hadron the Goldstone bosons are scattered off. In the SU(3) limit
the decomposition follows
\begin{eqnarray}
&& 8 \otimes \bar 3 = \bar 3\oplus 6 \oplus \overline{15}\,, \qquad \qquad \qquad \qquad \qquad \quad
8 \otimes 6= \bar 3\oplus 6 \oplus \overline{15} \oplus 24 \,.
\label{decomp-c}
\end{eqnarray}
For $8 \otimes \bar 3$ scattering the leading order chiral Lagrangian predicts
attraction in the anti-triplet and sextet sectors, but
repulsion for the anti-15plet.  For $8 \otimes 6$ scattering there is
attraction in the anti-triplet, sextet but also in the anti-15plet sectors. The interaction
is repulsive in the 24plet sectors. Chiral SU(3) symmetry predicts a hierarchy of
strengths with the strongest attraction in the triplet channels, which
is five  times as strong as the attraction in the anti-15plet. In the
sextet sector the attraction is reduced by a
factor $3/5$ only. It is also rewarding to compare the amount of
attraction in the anti-triplet channels as they result from the
reduction of $8 \otimes \bar 3 $ versus $8 \otimes 6$. Chiral
symmetry predicts stronger binding in the latter case. The amount
of attraction is larger by a factor $5/3$ as compared to the
former case

\begin{figure}[t]
\vspace{-0.5cm}
{\parbox{10.7cm}{\includegraphics[width=10.0cm,clip=true]{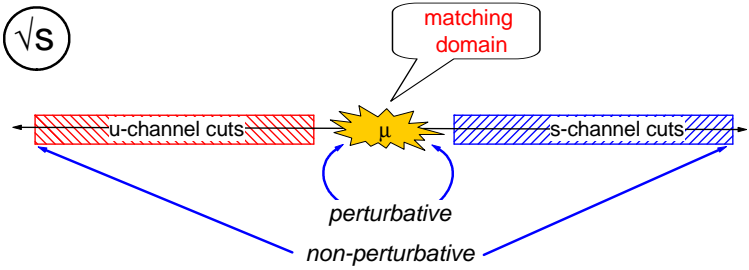}}}
{\parbox{5cm}{\vspace*{-1cm}
\caption{Typical s- and u-channel cut structure of the scattering amplitude in the complex
energy plane
$\sqrt{s}$.}}} \label{fig:1}
\end{figure}

Given all these interesting predictions, how do we translate this information into a spectrum of
resonances? Clearly, if the interaction of the Goldstone bosons with a hadron is sufficiently
attractive in some channel, one would expect the formation of a molecule. Unfortunately, it is a
priori ill-defined to insert the interaction vertex (\ref{WT-term}) into a coupled-channel
scattering equation, like the Bethe-Salpeter or Lippmann-Schwinger equation. The quasi-local
nature of the interaction gives rise to short-range divergencies. A renormalization program
can be set up in a perturbative approach defining unique answers at each order in the perturbation.
However, the formation of a molecule requires the summation of  an infinite number of diagrams.
This asks for novel tools. In a phenomenological approach one may introduce a cutoff to tame the
coupled-channel equation, however, the price is that the binding energy of the such molecule
depends sensitively on the precise value of the cutoff. More seriously, the introduction of a
cutoff in a meson-baryon system is at odds with the counting laws that permit a truncation of
the interaction in the first place. Rather, then giving a formal discussion of these
problems (see \cite{LK02,LK04-axial,HYP03}), an intuitive and physical argument how to
remedy this unsatisfactory situation is given.

\begin{figure}[t]
\vspace{-0.8cm}
\parbox{10.7cm}{\includegraphics[width=10.0cm,clip=true]{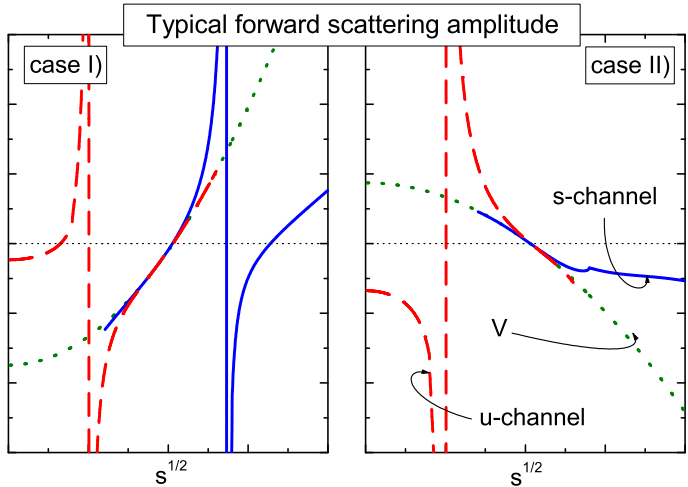}}
\parbox{5cm}{\vspace*{-1cm}
\caption{Typical cases of forward scattering amplitudes. The solid
(dashed) line
 shows the s-channel (u-channel)
unitarized scattering amplitude. The dotted lines represent the
amplitude evaluated at tree-level.}} \label{fig:2}
\end{figure}

The problem is illustrated in Fig. 1, which indicates that the physics of a
partial-wave scattering amplitude is perturbative in a small subthreshold energy domain only.
The amplitude is characterized by its s-channel and u-channel unitarity branch points. Additional
branch points implied by t-channel exchanges are suppressed in the discussion for simplicity.
At sufficiently large and small energies  the amplitude is necessarily non-perturbative
with possible resonance signals. If one feeds an interaction kernel into a coupled-channel scattering
equation the s-channel unitarity cut is constructed explicitly, i.e. the scattering amplitude
satisfies s-channel coupled-channel unitarity. However, unless an infinite sum of diagrams is
considered the so constructed scattering amplitude does not satisfy u-channel unitarity, i.e the
approximation to the u-channel cut is necessarily inadequate. Thus, this amplitude can be trusted
only at energies to right of $\mu$ in Fig. 1. In practical applications this is not really
an issue. The u-channel cut can be reconstructed uniquely by the s-channel cuts of the crossed
reactions. Since the scattering amplitude has a small window where it can be evaluated
in perturbation theory, an accurate scattering amplitude, applicable to the left and right of
$\mu$ in Fig. 1, can be constructed by gluing together s- and u-channel unitarized amplitudes at
some matching point $\mu$. As a result the full amplitude is crossing symmetric exactly
at energies accessible to experiments. The glued amplitude should be smooth and
not jump around at the matching point. This is not necessarily realized in phenomenological
coupled-channel computations, in which a cutoff or a form factor may severely destroy the
perturbative nature of the subthreshold amplitudes. In \cite{LK02} we suggested to take the
smooth matching as a constraint to be imposed on the unitarization scheme. This sifts out a
major part of the ambiguities inherent in unitarization schemes. The renormalization program is
carried out in the small and perturbative subthreshold window. Any residual scheme dependence is
systematically eliminated upon the inclusion of higher order contributions to the interaction kernel.

In Fig. \ref{fig:2} we demonstrate the quality of the
proposed matching procedure as applied for typical forward
scattering amplitudes \cite{LK04-axial}. The figure clearly illustrates a successful
matching of s-channel and u-channel iterated amplitudes at
subthreshold energies.

\section{S- and d-wave baryon resonances with zero charm}

\begin{figure}[t]
\begin{center}
\parbox{17.5cm}{\includegraphics[clip=true,width=17.5cm]{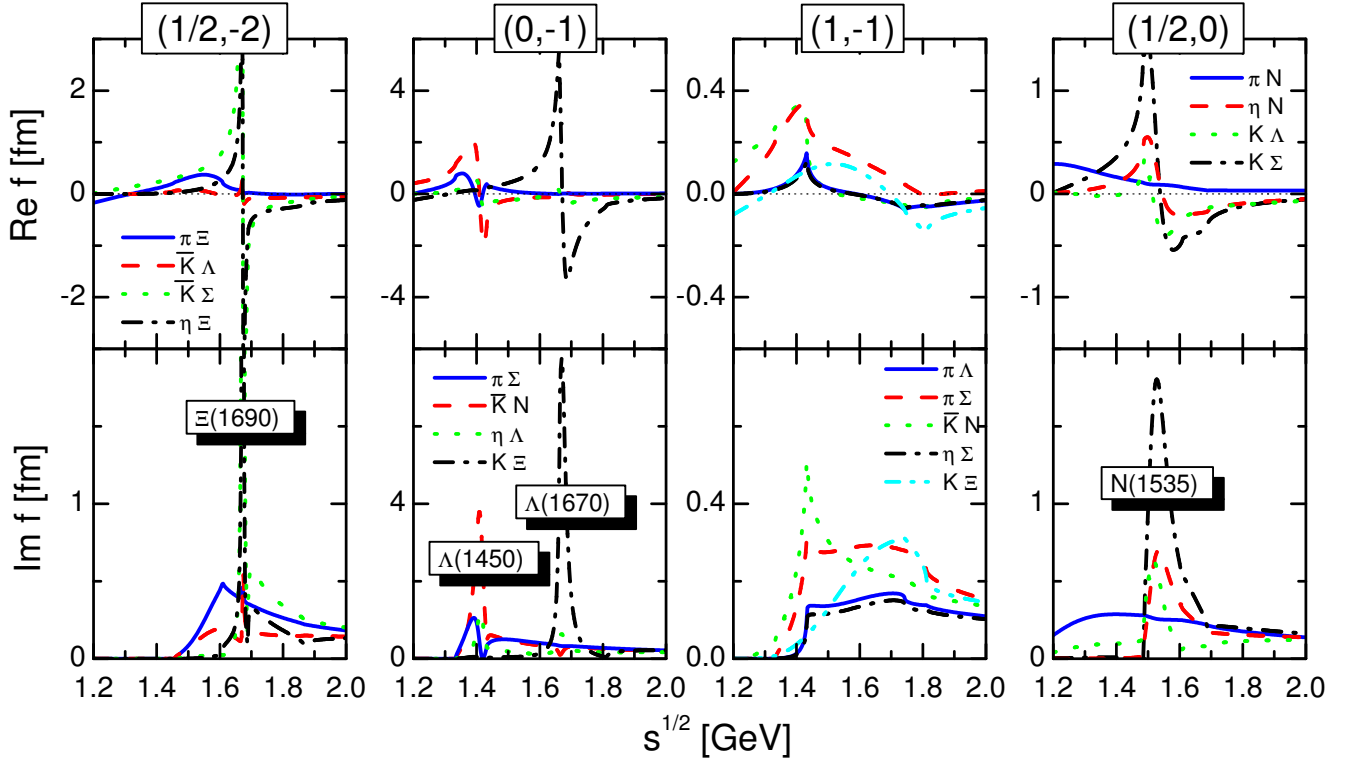}}
\end{center}
\caption{Baryon resonances as seen in
the s-wave scattering of Goldstone bosons off the baryon octet $N(939), \Lambda(1115), \Sigma(1195),
\Xi(1315) $. Shown are real and imaginary parts of partial-wave scattering amplitudes
with $J^P=\frac{1}{2}^-$ and $(I,S)=(1/2,-2),
(0,-1),(1,-1),(1/2,0)$.} \label{fig:3}
\end{figure}

We discuss the chiral excitations of the baryon octet and decuplet ground states of QCD.
The partial-wave scattering amplitudes of Fig. \ref{fig:3} show evidence for the formation of the $\Xi(1690)$,
$\Lambda(1405)$, $\Lambda(1670)$ and $N(1535)$ resonances close to their empirical masses. An
additional $(I,S)=(0,-1)$ state, mainly a SU(3) singlet \cite{Jido03,Granada}, can be found as a
complex pole in the scattering amplitude close to the pole implied by the $\Lambda(1405)$ resonance.
There is no convincing signal for a $(1,-1)$ resonances at this leading order calculation.
However, chiral corrections lead to a pronounced resonance signal in this sector
\cite{LK02,Granada} which should be identified with the $\Sigma(1750)$ resonance, the only
well established s-wave resonance in this sector. It couples strongly to the $K \,\Xi$ and
$\eta \,\Sigma$ states. The fact that a second resonance with $(\frac{1}{2},0)$ is not seen in
Fig. \ref{fig:3}, even though the `heavy' SU(3) limit suggests its existence
\cite{Granada,KL03}, we take as a confirmation of the phenomenological observation that the
$N(1650)$ resonance couples strongly to the $\omega \,N$ channel not considered
here \cite{LWF02}. For a more detailed discussion of the spectrum we refer to \cite{Granada}.

\begin{figure}[t]
\centerline{
\parbox{7.0cm}{\includegraphics[clip=true,width=7.0cm]{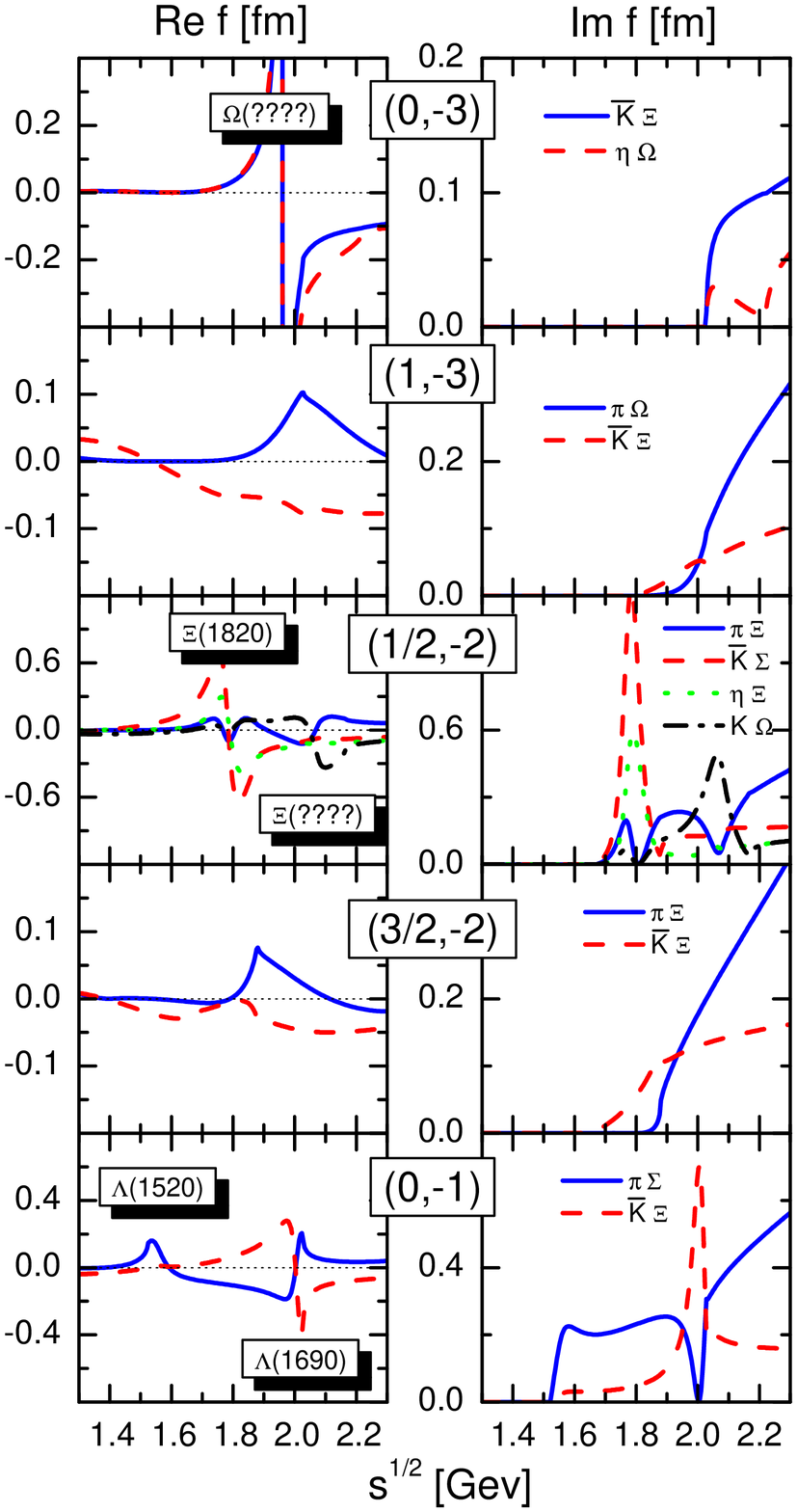}}
\quad
\parbox{7.2cm}{\includegraphics[clip=true,width=7.2cm]{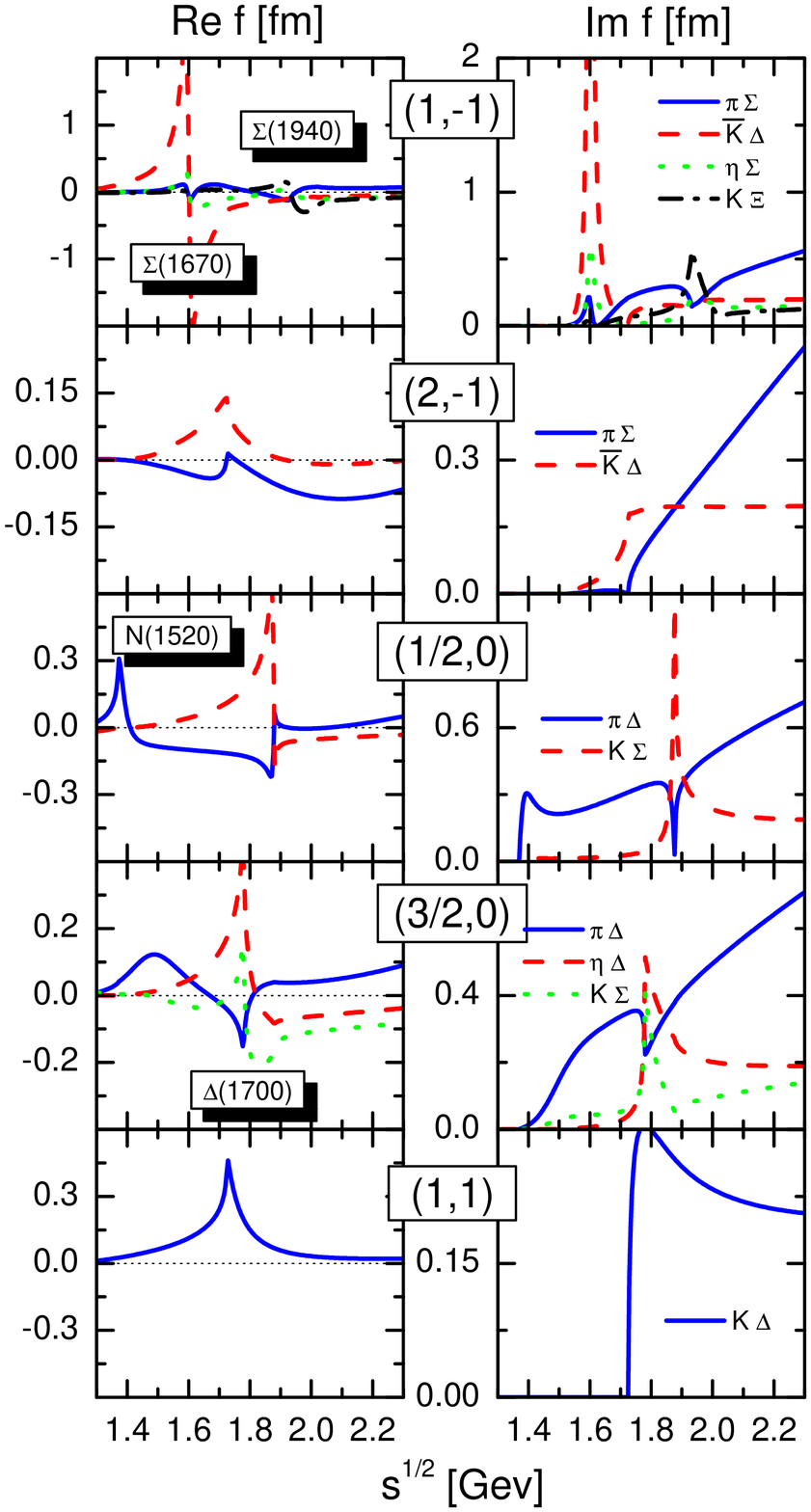}}
}
\caption{Baryon resonances with $J^P=\frac{3}{2}^-$ as formed in
the s-wave scattering of Goldstone bosons off the baryon decuplet $\Delta(1232), \Sigma(1385), \Xi(1530),
\Omega (1672)$. Shown are real and imaginary parts of partial-wave scattering amplitudes.} \label{fig:4}
\end{figure}

We turn to the d-wave resonance spectrum, which can be read off Fig. \ref{fig:4}. Amplitudes,
that describe the s-wave scattering of Goldstone bosons off the decuplet states are shown.
All sectors are included for which chiral symmetry predicts an attractive interaction.
One may expect that chiral dynamics does not make firm
predictions for d-wave resonances since the meson baryon-octet
interaction in the relevant channels probes a set of
counter terms presently unknown. However, this is not
necessarily so. Since a d-wave baryon resonance couples to s-wave
meson baryon-decuplet states chiral symmetry is quite predictive
for such resonances, under the assumption that the
latter channels are dominant. This is in full analogy to an
analysis of the s-wave resonances  \cite{Wyld,Dalitz,sw88,Ji01,grkl,Jido03,Granada}
that neglects the effect of the contribution of d-wave meson baryon-decuplet
states. The empirical observation that the d-wave resonances
$N(1520)$, $N(1700)$ and $\Delta (1700)$ have large branching
fractions ($> 50 \% $) into the inelastic $N \pi \pi$ channel, even
though the elastic $\pi N$ channel is favored by phase space,
supports this assumption.
It is a stunning success of chiral coupled-channel dynamics that it recovers the
four star hyperon resonances $\Xi(1820)$, $\Lambda(1520)$, $\Sigma (1670)$ with masses quite
close to the empirical values. The nucleon and isobar resonances $N(1520)$ and $\Delta (1700)$
also present in Fig. \ref{fig:4}, are predicted with less accuracy. The important result here is
the fact that these resonances are generated at all. Fully realistic results should not be
expected already in this leading order calculation. For instance, chiral correction
terms provide a d-wave $\pi \Delta$ component of the $N(1520)$. Moreover, from phenomenological
analysis one anticipates the importance of additional inelastic channels involving
vector mesons (see e.g. \cite{LWF02}). We continue with the bound state signal in the $(0,-3)$
amplitudes at mass 1950 MeV. Such a state has so far not been observed but is associated with a
decuplet resonance \cite{Schat}. Further states belonging to the decuplet are seen in the
$(\frac{1}{2},-2)$ and $(1,-1)$ amplitudes at masses 2100 MeV and 1920 MeV. The latter state can
be identified with the three star $\Xi (1940)$ resonance. Finally we point out that the
$(0,-1)$ amplitudes show signals of two resonance states consistent with the existence of the
four star resonance $\Lambda(1520)$ and $\Lambda(1690)$. It appears that the SU(3) symmetry
breaking pattern generates the `missing' state in this particular sector by promoting the weak
attraction of the 27plet. For completeness we include also the exotic sectors
$(1,-3),(3/2,-2),(2,-1),(1,1)$ part of the
27plet. Since chiral dynamics predicts weak attraction only, no clear resonance signals are
seen in the amplitudes. It remains an open issue whether correction terms may conspire to
increase the amount of attraction in some selected channels leading to unambiguous exotic signals.
Subsequently our results have been confirmed in \cite{Sarkar:2004jh}.

\begin{figure}[t]
\centerline{
\parbox{7.5cm}{\includegraphics[clip=true,width=7.5cm]{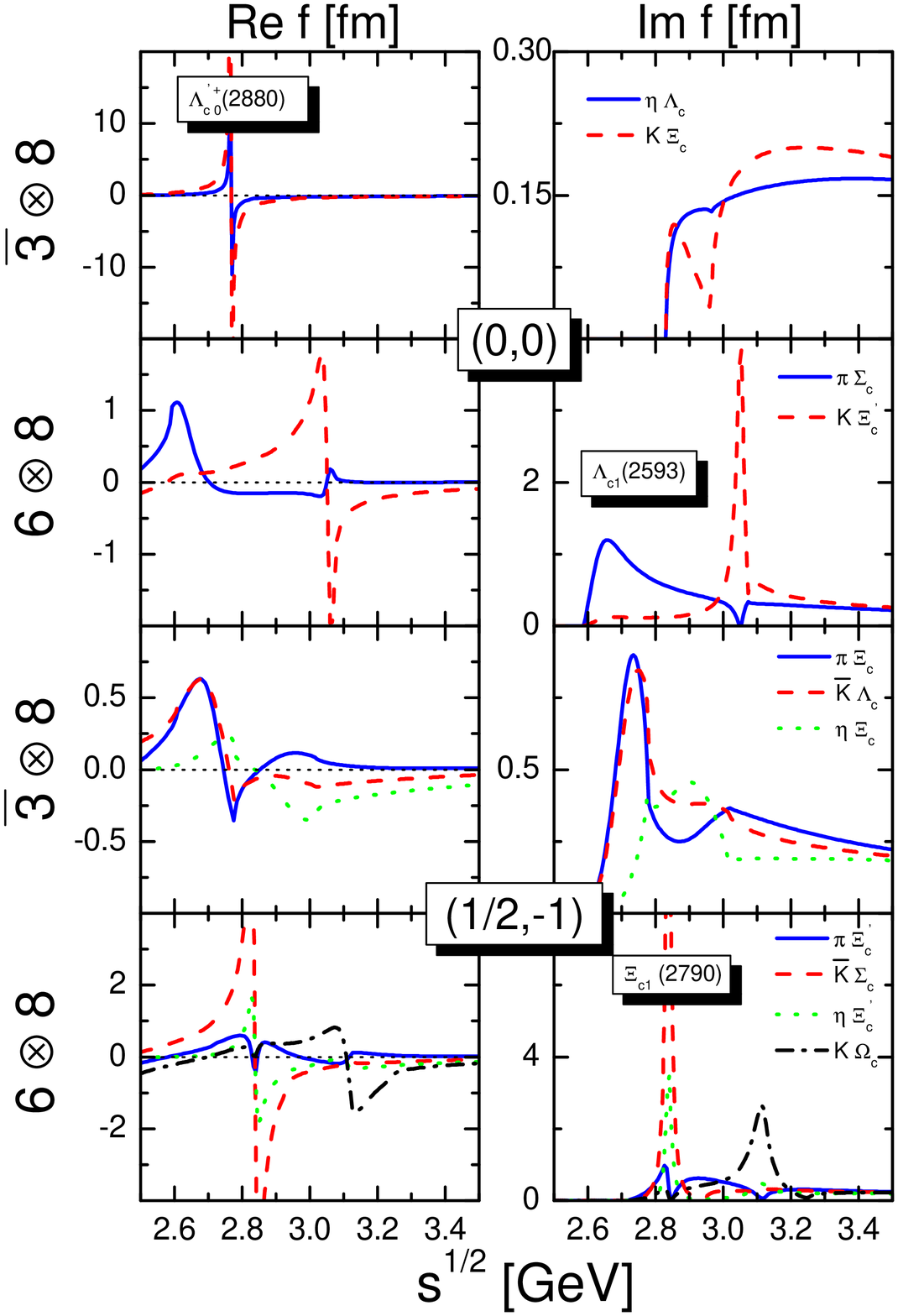}}
\hfill
\parbox{7.5cm}{\includegraphics[clip=true,width=7.5cm]{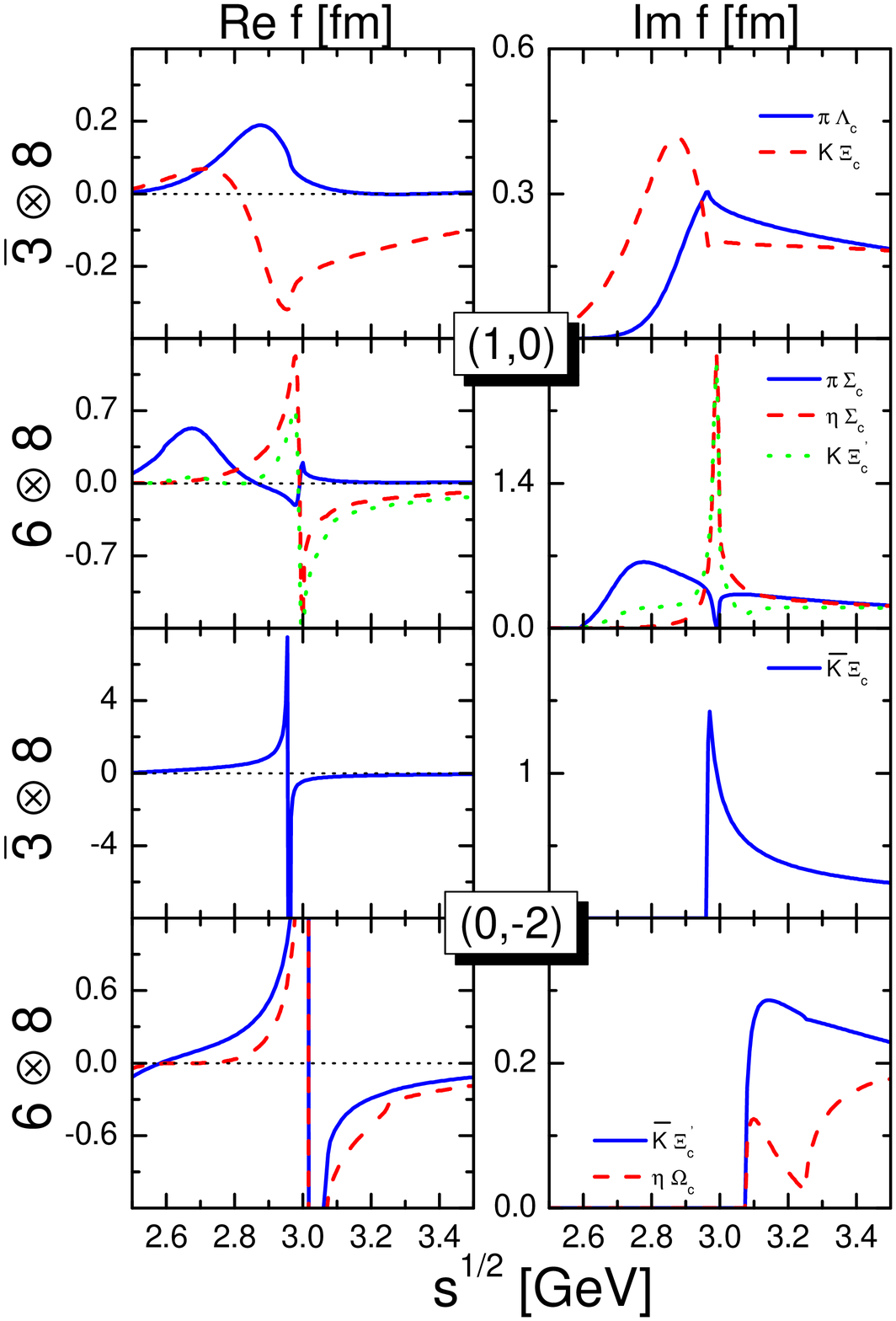}}
}
\caption{Open-charm baryon resonances as seen in
the scattering of Goldstone bosons off anti-triplet $\Lambda_c(2284), \Xi_c(2470)$ and
sextet $\Sigma_c(2453),\Xi'_c(2580),\Omega_c(2697)$ baryons.
Shown are real and imaginary parts of partial-wave scattering amplitudes
with $J^P=\frac{1}{2}^-$  and $(I,S)=(0,0),(1/2,-1),
(1,0),(0,-2)$.} \label{fig:5}
\end{figure}

\section{S- and d-wave baryon resonances with open charm}

\begin{figure}[t]
\centerline{
\parbox{7.0cm}{\includegraphics[clip=true,width=7.0cm]{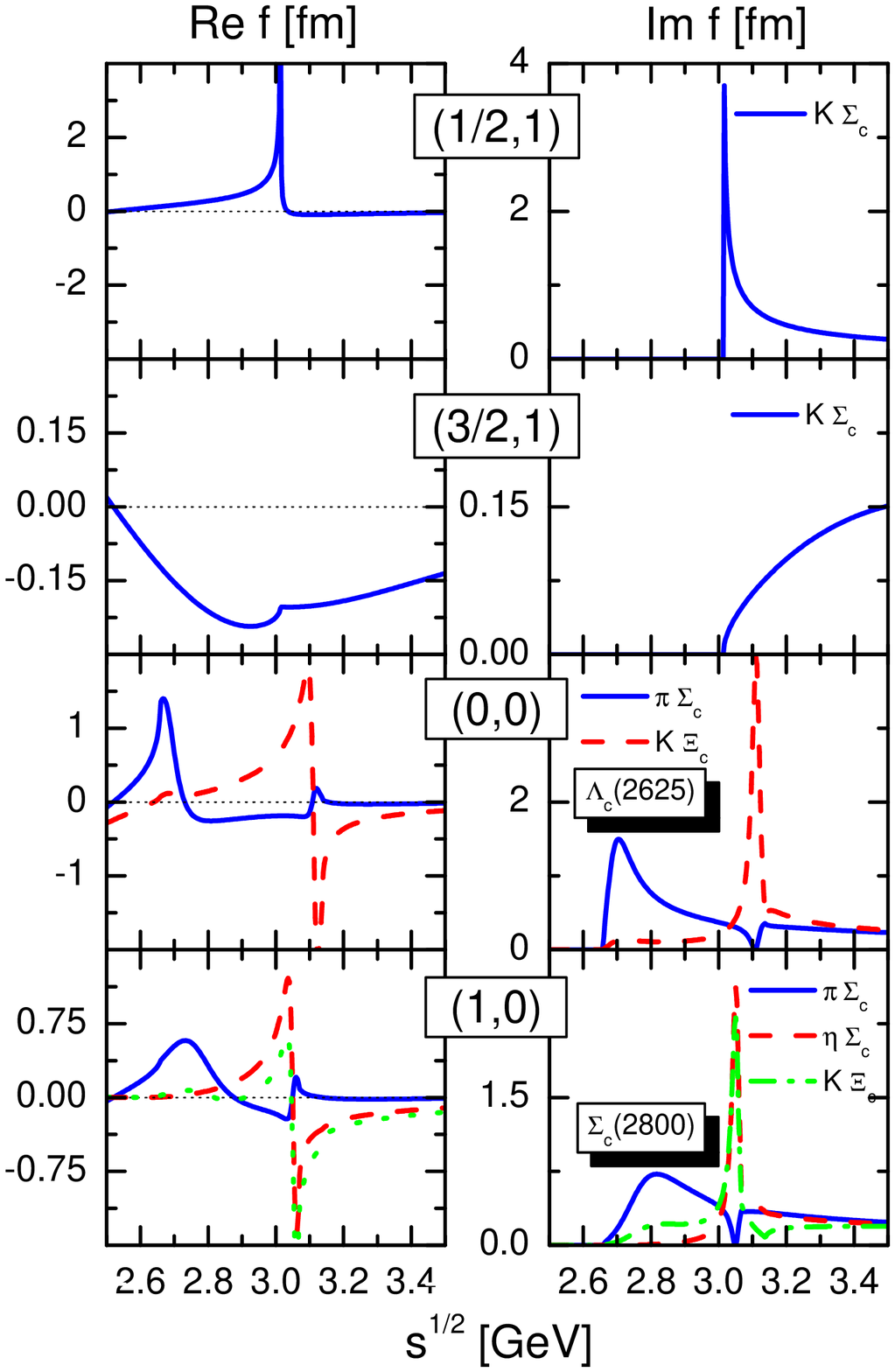}}
\quad
\parbox{7.0cm}{\includegraphics[clip=true,width=7.0cm]{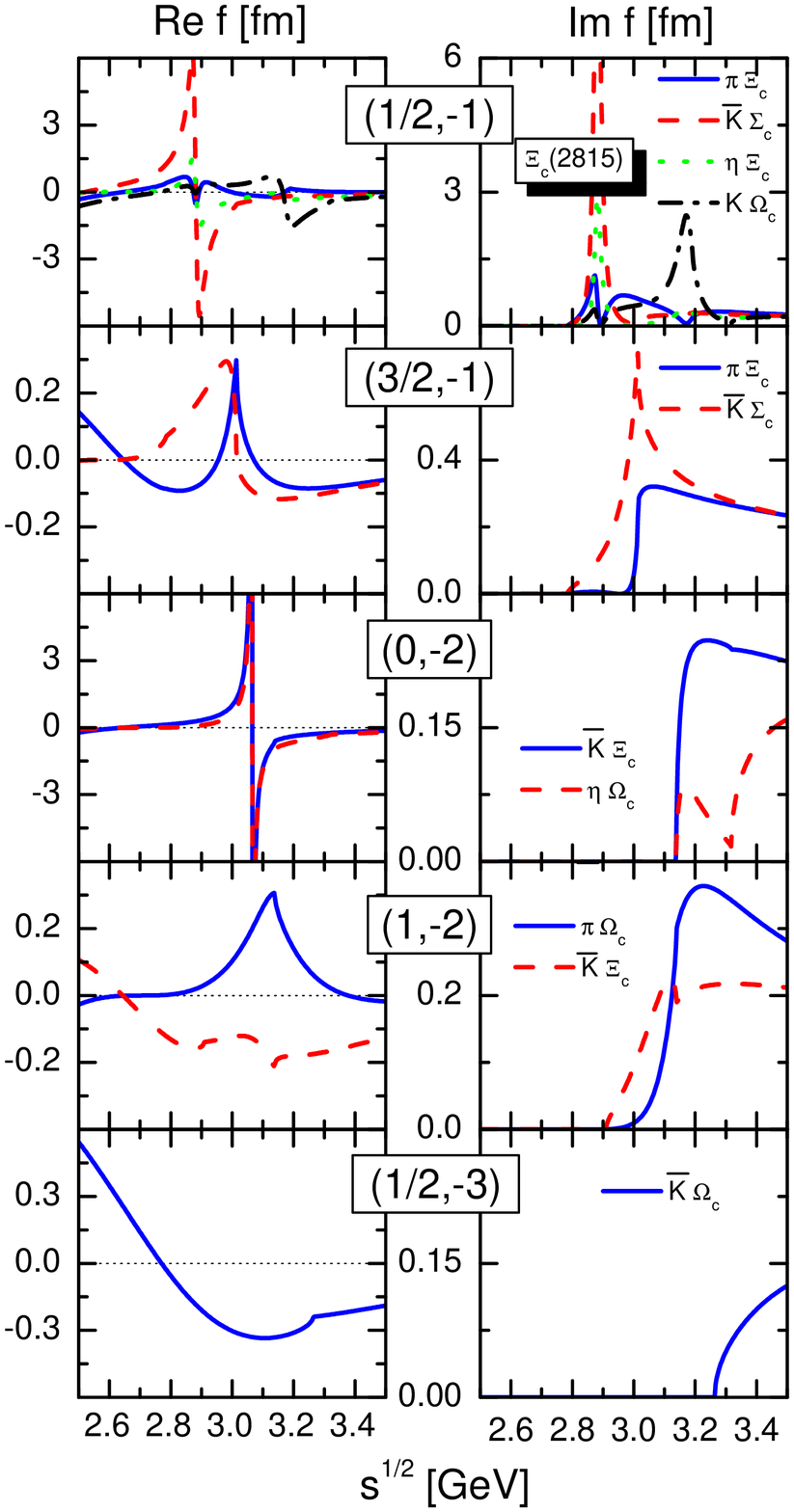}}
}
\caption{Open-charm baryon resonances with $J^P=\frac{3}{2}^-$ as seen in
the scattering of Goldstone bosons off the
sextet $\Sigma_c(2520),\Xi_c(2644),\Omega_c(2770)$ baryons.
Shown are real and imaginary parts of partial-wave scattering amplitudes.} \label{fig:6}
\end{figure}

We turn to the chiral excitations of open-charm baryons.
At present there is little known empirically
about the open-charm baryon resonance spectrum. Only three s-wave resonances
$\Lambda_{c}(2593), \Lambda_{c}(2880)$ and $\Xi_{c}(2790)$ were
discovered so far \cite{PDG04}. Two d-wave resonances $\Lambda_c(2625)$ and
$\Xi_c(2815)$ are acknowledged by the particle data group \cite{PDG04}.
An additional isospin triplet state $\Sigma_c(2800)$ was announced recently by the BELLE
collaboration \cite{BELLE-Sigma}, which most likely carries $J^P=\frac{3}{2}^-$
quantum numbers also.

We confront the empirical spectrum with the spectrum of chiral molecules.
Let us first discuss the chiral excitations of the $J^P=\frac{1}{2}^+$ states, which form an
anti-triplet and sextet. The relevant scattering amplitudes are collected in Fig. \ref{fig:5}.
Amplitudes with identical quantum numbers that describe the scattering of Goldstone bosons off the
anti-triplet and sextet states are grouped on top of each other. This is instructive
since at subleading order in the chiral expansion such amplitudes start to couple.

Consider the chiral excitations of the triplet states manifest in the first and third
row of Fig. \ref{fig:5}. According to our discussion those states form a strongly
bound anti-triplet and a weakly bound sextet. The $(0,0)$ resonance is identified with
the $\Lambda_c(2880)$ being a member of the anti-triplet. It is in fact a bound state
at this leading order computation. The isospin doublet of that multiplet, a $(1/2,-1)$ state,
is a much broader object so far unobserved. The chiral sextet excitations of the anti-triplet
is visible in the $(1,0),(1/2,-1)$ and $(0,-2)$ sectors. Most spectacular
would be the exotic and bound $\bar K\,\Xi_c(2470)$ system, if it survives chiral correction terms.

An even richer spectrum is formed by the chiral excitations of the sextet
states as shown in the second and fourth row of Fig. \ref{fig:5}.
Chiral SU(3) symmetry predicts attraction with decreasing strength in the triplet,
sextet and anti-15plet. The triplet states are identified with the $\Lambda_c(2593)$
and $\Xi_c(2790)$ resonances. The sextet manifests itself most clearly as a bound
$\bar K\,\Xi_c'(2580)$ system. The existence of the anti-15plet is clearly seen in the $(0,0),(1,0)$
and $(1/2,-1)$ sectors where it leads to narrow resonance structures around 3 GeV.
Additional signals of the anti-15plet in the exotic sectors $(1/2,1),(3/2,-1)$ and $(1,-2)$
are weaker and are not shown in Fig. \ref{fig:5}. They can, however, be inferred from Fig.
\ref{fig:6}.

We turn to the d-wave spectrum of open charm baryons. It is induced by the s-wave scattering
of the Goldstone bosons off the sextet $J^P=\frac{3}{2}^+$ ground states. Here we assume a
value for the $\Omega_c(2770)$ mass, so far unknown, which is suggested by quark
model calculations \cite{Isgur} and lattice simulations \cite{Mathur:2002ce}. The scattering
amplitudes are shown in Fig. \ref{fig:6} for all sectors in which chiral dynamics predicts
attraction. The pattern is analogous to the already discussed spectrum arising from the
scattering of the Goldstone bosons off the sextet of $\frac{1}{2}^+$ ground states. This
is an immediate consequence of the identical interaction strengths of the two systems.
So far only three states are observed experimentally $\Lambda_c(2625), \Sigma_c(2800)$
and $\Xi_c(2815)$. The masses are rather well reproduced, given the fact that the
amplitudes are computed at the parameter-free
leading order. We underestimate the binding for the $\Lambda_c(2625)$ and $\Xi_c(2815)$
by about 75 MeV. Whereas the width of the $\Sigma_c(2800)$ is quite consistent with the
recent BELLE measurement we overestimate the width of the $\Xi_c(2815)$
and $\Lambda_c(2656)$. Note that these states are somewhat underbound. Allowing
for small attractive correction terms will pull
down these states giving them a smaller width within reach of their measured values.
Besides the anti-triplet excitations, chiral dynamics predicts again a sextet and anti-15plet
of excitations, however, with reduced binding strength. Most spectacular is the prediction that
the $K$ is bound at the $\Sigma_c(2520)$ and the
$\bar K$ at the $\Xi_c(2644)$. The former is rather weakly bound,
if at all, being a member of the anti-15plet. The latter enjoys a binding energy of about 40 MeV
being a member of the sextet. Since it couples equally strong to the $\eta\, \Omega_c(2774)$
state a derivation of its electromagnetic properties requires a coupled-channel study. Striking
evidence for the existence of the anti-15plet are narrow resonance
structures in the $(0,0)$ and $(1,0)$ sectors around 3.1 GeV. These resonances couple strongly to
the $K \,\Xi_c(2664)$ state. A further somewhat broader anti-15plet state with $(1/2,-1)$
quantum numbers is predicted at around 3.2 GeV. This state is dominated by its
$K \,\Omega_c(2774)$ component.

\section{Summary and outlook}

In  this talk we reviewed the spectrum of chiral excitations of the large-$N_c$
ground state baryons. The leading order term of the chiral Lagrangian predicts a rich
spectrum of molecules in astounding agreement with the empirical resonance pattern. No adjustable
parameters enter the spectrum. It is determined by the well known value of the pion decay constant.
The properties of the chiral excitations depend crucially on the values of the current quark masses
of QCD. An SU(3) limit where the pion mass is as heavy as the empirical kaon mass leads to a
bound state spectrum, typically. The SU(3) limit where the kaon mass is as light as the empirical
pion mass implies the disappearance of the resonance signals. This is a striking prediction of
chiral dynamics which should be verified by unquenched lattice QCD simulations. Besides
reproducing well-established states new, so far unobserved,
exotic multiplets are foreseen. New and unpublished results concern the d-wave spectrum of
open charm baryons. The recently announced $\Sigma_c(2800)$ state of the BELLE collaboration
is interpreted as a chiral molecule. The large width of
that state is naturally explained. Similarly, the $\Lambda_c(2625)$ and $\Xi_c(2815)$ states
are well understood in terms of chiral coupled-channel dynamics.
The results support the radical conjecture of the authors that
meson and baryon resonances that do not belong to the
large-$N_c$ ground state of QCD should be viewed as hadronic molecular states.
To further substantiate these findings extensive and refined coupled-channel computations
are necessary that will reveal the resonance spectrum of QCD more quantitatively.

\vskip0.3cm
{\bfseries{Acknowledgments}}
\vskip0.2cm
The work of E.E.K.\ was supported in part by the US Department of Energy
under contract No.\ DE-FG02-87ER40328.

\end{document}